# The Missing Link: Exploring the Relationship Between Transformational Leadership and Change in team members in Construction

*[1]Mubarak Reme Ibrahim

[1]Department of Building, Faculty of Environmental Sciences, Baze University Abuja, Nigeria

*Corresponding author: Mubarak Reme Ibrahim mubarakibrahim0607@gmail.com 07063560817

**Abstract**

The archetype of leadership serves as a crucial link between leaders and their followers, yet many studies on leadership within the construction industry have neglected to examine this vital connection. This study aimed to investigate how transformational leadership affects team processes, mediated by change in team members. The study had four objectives: to evaluate the relationship between transformational leadership and team processes, to assess the relationship between transformational leadership and change in team members, to examine the relationship between change team members and team processes, and to establish the mediating role of change in team members. A self-administered questionnaire was distributed to construction project team members in Abuja and Kaduna, and statistical analysis using path modeling in Smart PLS 3 software revealed a significant positive relationship between transformational leadership and team processes, transformational leadership and change in team members, changes in team members and team processes, and changes in team members mediating the relationship between transformational leadership and team processes. Future studies should consider cultural differences.

## 1.0 Introduction

Over the past decade, leadership research has focused on assessing the direct impact of leadership on organisational outcomes such as performance (Anantatmula, 2010; Larsson et al., 2015; Turner & Müller, 2005), organisational citizenship behaviour (Gilmore et al., 2013; López-Domínguez et al., 2013; Öztürk & Ay, 2015), job satisfaction (Gaviria-Rivera & Lopez-Zapata, 2019), organisational commitment, and individual and team performance and creativity (A. Ali et al., 2020; Baskoro, 2021; Khairuddin et al., 2021; Maqsoom et al., 2022). During this period, the emphasis of leadership research has moved from transactional to transformational archetypes of leadership (Akhavan Tabassi & Hassan Abu Bakar, 2010; Anwar, 2016; Oswald et al., n.d.; Y. Zhang et al., 2018). Transformational leadership has been described as a leadership style that motivates and inspires followers to achieve their goals and objectives beyond their self-interests (Akhavan Tabassi & Hassan Abu Bakar, 2010; Avolio et al., 1991; Avolio & Bass, 2001; Bass, 1999; Bass & Avolio, 1993, 1994; Den Hartog et al., 1997; Jung & Sosik, 2002).

Although there have been empirical research studies that have investigated team processes as an outcome variable and mediator variable with transformational leadership (Aga et al., 2016; Akhavan Tabassi et al., 2014; H. Ali et al., 2021; Dionne et al., 2004; Gaviria-Rivera & Lopez-Zapata, 2019; Lehmann-Willenbrock et al., 2015; Nauman et al., 2022; Tabassi et al., 2017), the association between these constructs has not been established with mediating mechanism of how transformational leadership influenced change in team members in terms of their attitude, motivation, and perception toward work.

However, little research has been conducted on the impact of transformational leadership on team processes when mediated by change in team members' attitude, motivation, and perception toward the task. Team processes are the cognitive, verbal, and behavioural activities that teams use to achieve their objectives. These processes include communication, collaboration, and conflict management. They are crucial for the achievement of team goals, and effective team processes contribute to the team's overall success. There have been few studies focused on the contribution of leadership behaviour on team processes and investigated the relationship between transformational leadership and team processes. Several studies have suggested that transformational leadership is positively associated with team processes. This suggests that project managers who demonstrate transformational leadership behaviours may positively impact team processes. However, few studies have focused on the mediating mechanisms that explain the relationship between transformational leadership and team processes. Research has suggested that team building and leader-member exchange can mediate the relationship between transformational leadership and project success.

The current study aims to extend the existing transformational leadership team processes models by assuming that transformational leadership behaviour from project managers influences change in team members' attitude, motivation, and perception towards the task, thereby mediating the relationship between transformational leadership and team processes. The study will examine whether transformational leadership behaviour leads to change in team members' attitude, motivation, and perception toward the task, and whether these changes mediate the relationship between transformational leadership and team processes.

## 2.0 Literature Review

In this section, we will outline the theoretical basis for the three key components of the study: team processes, transformational leadership, and change in change in change in team members.

## 2.1 Team Processes

Most team effectiveness models, including those developed by (Guzzo & Shea, 1992), (Gist et al., 1987), and (Hackman, 1983; Tannenbaum et al., 1992), place a central focus on team processes. One model that has been influential in organisational studies for several decades is McGrath's (1964) input-process-outcome (IPO) model. Inputs represent the antecedent factors that facilitate or hinder team member interaction, including individual member characteristics, team-level factors, and organisational and contextual factors. Outcomes refer to the valued results, such as timely completion, efficient resource utilization, and high-quality outputs. Team processes serve as the linking mechanism between inputs and outcomes and refer to how change in change in change in team members interact toward the project goal. Team processes can be divided into three classifications proposed by (Marks et al., 2001): transition (mission analysis, planning, goal specification, and strategy formulation), action (task accomplishment, progress monitoring, coordination, and self-monitoring/support), and interpersonal (conflict management, motivation and confidence-building, and affect management). Traditionally, team processes were categorized as either "taskwork" or "teamwork," where task work refers to roles individuals must fulfill to accomplish the team's task, while teamwork describes the interaction among change in change in change in team members.

## 2.1.1 Transition Process

(Marks et al., 2001) proposed that team processes occur periodically, with transition processes setting the stage for subsequent actions. While there is limited empirical research on transition processes, some studies have shed light on their importance. For example, (Chong, 2007; Janicik & Bartel, 2003) found that planning helped establish norms related to time management, which were positively associated with performance. Similarly, (Chiu et al., 2016; Hiller et al., 2006; Motowidlo & Van Scotter, 1994; Riggio et al., 2003; Z. Zhang et al., 2012) discovered that shared leadership behaviours, such as planning and organizing, significantly predicted supervisor-rated team performance.

Teams that engage in transition processes tend to perform better than those that do not. In particular, these teams are more likely to focus on coordination issues, which ultimately leads to better performance (LePine et al., 2008; Olson et al., 1992). (DeChurch & Haas, 2008; Martin, 2011; J. E. Mathieu & Schulze, 2006) used an episodic model of team processes and argued that dynamic planning, including contingency and reactive planning, was significantly related to performance in their study of a project. Furthermore, (Mathieu et al., 2008) demonstrated how initial team activities, such as the quality of deliberate performance plans and team charters, were positively associated with the forms of team performance displayed over time. Putting together, these studies suggest that transition processes are a critical component of effective teamwork. By setting the stage for subsequent actions and helping establish norms related to time management and coordination, transition processes can contribute to better team performance over time.

### 2.1.2 Action processes

While transition processes have not been the primary focus of research in teams, action processes such as communication and coordination have been well-studied and found to play key roles in team performance (LePine et al., 2008). Research has shown that effective communication and coordination within teams are essential to successful problem management and work group effectiveness (Ellis et al., 2005; Hagemann & Kluge, 2017; Hung, 2013; Mishra et al., 2012; Piccoli et al., 2004). The quality of communication can vary greatly between teams depending on factors such as the frequency of exchanges, formalization, and communication structure, all of which can impact the quality of action processes (Hoegl & Gemuenden, 2001).
Numerous studies have shown that high levels of (Daniel et al., 2013; Espinosa et al., 2006; Wu et al., 2017; Yang et al., 2012). Additionally, coordination has been identified as a critical component for project success, as it allows teams to effectively respond to problems that arise during task progression (Alaloul et al., 2016; Bose, 2002; Ika et al., 2012; K. N. Jha & Iyer, 2006). Effective coordination can lead to better problem management and higher levels of performance. Participation and reward systems are also important factors in team performance. (De Dreu & West, 2001) found that team member participation, particularly in the presence of minority opposition, can increase team creativity. Similarly, (Johnson et al., 2006) reward system can affect their level of information sharing and as a result, shape the speed and accuracy of their decision-making. Additionally, backup behaviours have been found to be significantly related to decision-making performance (Connor & Becker, 2003).

### 2.1.3 Interpersonal processes

The interpersonal process component of team dynamics encompasses conflict, motivation, confidence building, and affect. While all these components have been the focus of research, conflict has received the most attention. Studies have shown that both relationship and task conflicts have positive and negative correlations with team performance and member satisfaction (De Dreu, 2006; De Dreu & Van Vianen, 2001; De Dreu & Weingart, 2003; DeChurch & Marks, 2001; Jiang et al., 2013; O'Neill et al., 2013). Task conflict has been found to be influenced by informational diversity and has a significant impact on performance. Relationship conflict, on the other hand, has been found to mediate the relationship between ambient sexual hostility and team financial performance (Bradley-Geist et al., 2015).

Apart from conflict, research has also explored other interpersonal processes that impact team dynamics. Feedback has been found to have a significant impact on motivation, interpersonal trust, and overall performance in virtual teams (Geister et al., 2006; Hertel et al., 2005; Jarvenpaa et al., 1998). A combined metric of interpersonal processes has been positively associated with performance, and it provides the cross-level mechanism by which team-level resistance to empowerment climate is linked to individual-level satisfaction. Temporal components of team and task have also been found to be crucial to the effect of interpersonal mediations on team performance. Previous studies have been carried out with teams that briefly completed contrived tasks and were then disbanded, but more recent studies have shown that interpersonal processes significantly relate to team performance when teams are involved in longer-term tasks (Bradley et al., 2003; De Guinea et al., 2012; Druskat & Kayes, 2000; J. E. Mathieu & Rapp, 2009).

A meta-analysis on team conflict has shown that interpersonal processes, particularly conflict, perform a key role in project success (Chan et al., 2001; Wu et al., 2017). Task conflicts have been found to occur without relationship conflicts and are less likely to be emotional, escalate, or impair group performance. Relationship conflicts are less negatively linked to group performance in project teams, which suggests an interesting area for future research. The definition of project performance or project team performance should also be considered, and flushing out different types of performance metrics may uncover unique effects and help explain the influence of conflict on project performance.

## 2.2 Transformational Leadership

Leadership has been a topic of academic study for many decades, but there is a lack of empirical work on leadership in project management contexts. Although full-range leadership theory is a widely recognized theory of leadership, it encompasses transformational, transactional, and laissez-faire styles (Antonakis, 2001; Bass & Avolio, 1994; Deluga, 1990; Eagly et al., 2003). For the purpose of this discussion, the focus is on transformational leadership since studies have shown its high relevance for project-oriented organisations. In the literature, there is general agreement on four dimensions that make up transformational leadership: idealized influence, intellectual stimulation, inspirational motivation, and individualized consideration. Idealized influence is behaviour that arouses strong emotions and identification with the leader. Inspirational motivation is shown when a leader conveys a vision that is appealing and inspiring to subordinates and provides them with challenging assignments and increased expectations. Intellectual stimulation is behaviour that increases followers' awareness of problems and influences them to develop innovative and creative approaches to solving them. Individualized consideration includes providing support, encouragement, and coaching to followers.

These dimensions of transformational leadership have been identified as important for project-oriented organisations, and studies have shown that leaders who exhibit these behaviours are more likely to succeed in such contexts (Turner & Müller, 2005). However, there is still much to be explored in terms of how these leadership styles can be effectively applied in project management contexts (Den Hartog et al., 1997). Further research is needed to understand how transformational leadership can be harnessed to improve project outcomes and how it may interact with other important factors in project success (Brock & Von Wangenheim, 2019; Hassan et al., 2017; Shafi et al., 2020).

## 2.3 Change in team members due to the Influence of Transformation Leadership behaviour

### 2.3.1 Attitude

Attitudes are reflections of how individuals perceive their environment, and they can be assessed through statements about objects, people, or events. For example, when someone says "I like my job," they are expressing their attitude about their work. According to studies, attitudes generally have three dimensions: cognition, affect, and behaviour (Breckler, 1984). The cognitive part explains the way things are, such as "My pay is low," and sets the stage for the more critical aspect of an attitude. The affective part is the feeling component of an attitude, and it's revealed in statements like "I am angry over how little I'm paid." This feeling can lead to behavioural outputs, as the behavioural dimension of an attitude explains an intention to behave in a particular way toward something, such as "I'm going to look for another job that pays better." While earlier studies on attitudes argued that they were positively related to behaviour, a review of literature in the late 1960s criticized this assumed effect of attitudes on behaviour (Wicker, 1969).

Research has mostly established that people seek consistency among their attitudes and between their attitudes and their behaviour. They may either change their attitudes or behaviour or create a rationalization for the discrepancy (Fabrigar et al., 2006). The most powerful moderators of the attitudes relationship are the significance of the attitude, its correspondence to behaviour, its accessibility, the presence of social pressures, and whether a person has direct experience with the attitude. Attitudes that reflect fundamental values, self-interest, or identification with individuals or groups we value tend to show a strong association with our behaviour. Particularly, attitudes tend to predict specific behaviours, whereas general attitudes tend to best predict general behaviours. For example, asking someone about their intention to stay with an organisation for the next six months is likely to better predict turnover for that person than asking about their overall job satisfaction. On the other hand, overall job satisfaction would better predict a general behaviour, such as whether the individual was engaged in their work or motivated to contribute to their organisation (Harrison et al., 2006).

The attitude-behaviour relationship is likely to be much stronger if an attitude refers to something with which we have direct personal experience. Discrepancies between attitudes and behaviour tend to happen when social pressures to behave in a specific manner hold exceptional power, as in most situations. Ultimately, attitudes reflect how we perceive our environment, and understanding their dimensions can help predict our behaviour in different contexts (Klöckner, 2013).

### 2.3.2 Perception

Perception is the process by which individuals form and interpret their sensory impressions in order to give meaning to their surroundings (Ingold, 2002). However, what we perceive can be considerably different from objective reality. Perception can be influenced by a number of factors that can reside in the perceiver, in the object or target being perceived, or in the context of the situation in which the perception is made (Nisbett & Miyamoto, 2005). Personal characteristics such as attitudes, personality, motives, interests, past experiences, and expectations can deeply influence interpretation (Crewson, 1997). Characteristics of the target can also affect perception. People are more likely to notice loud people or extremely attractive or unattractive individuals in a group. The relationship of a target to its background can also influence perception.

The time, location, light, heat, and situational factors also matter as context influences perception. Attribution theory explains how people are judged differently, depending on the meaning of the attribute given to a behaviour (Kelley & Michela, 1980). Attribution depends on three factors: distinctiveness, consensus, and consistency. The halo effect operates when we draw a general impression about an individual on the basis of a single characteristic such as intelligence, sociability, or appearance. Contrast effect can distort perceptions as our reaction is influenced by other persons we have recently encountered. Stereotyping occurs when we judge someone on the basis of our perception of the group to which he or she belongs (Weiner, 2012).

### 2.3.3 Motivation

Motivation is a complex and fascinating area of study, as it touches on the very heart of what drives human behaviour (Weiner, 2012). Understanding what motivates people can be incredibly powerful, whether it's in the context of individual achievement or team success. By exploring the underlying factors that influence our choices and actions, we can gain insight into how to create environments that promote positive motivation and encourage people to perform at their best (Deci & Ryan, 2013). The concept of motivation is both simple and complex. At its simplest level, it is concerned with why people do what they do. However, as we delve deeper into the topic, we discover that motivation is a multifaceted and dynamic process that involves a range of internal and external factors (Lang, 1995). These can include things like personal goals, social influences, past experiences, and the perception of reward or punishment.

One of the key factors that underpins motivational theory is the notion of individuality. People are unique, with their own set of beliefs, values, and personality traits that shape their behaviour. As such, it is important to recognize that what motivates one person may not necessarily motivate another. By understanding the specific needs and desires of individuals, we can tailor our approach to motivation and create a more effective and engaging environment (Dutton et al., 1994).

Another important aspect of motivation is the idea of intentionality. People are not passive recipients of motivational factors, but rather active participants who choose to engage in certain behaviours based on their own desires and goals (Crant, 2000). Of course, motivation is not a static process. It is influenced by a range of internal and external factors that can change over time. This is why it is important to recognize that motivation is multifaceted, and that it involves both arousal (what gets people activated) and the direction or choice of behaviour (the force of an individual to engage in desired behaviour). By understanding these underlying factors, we can predict behaviour and create an environment that supports positive motivation.

### 3.0 Research Hypotheses

The section discusses the hypotheses of a study that explores the relationship between transformational leadership, change in change in change in change in team members as a result of transformational leadership behaviour from the project manager, and team processes. It suggests that, change in change in change in change in team members as a result of transformational leadership behaviour mediates the relationship between transformational leadership and team processes.

### 3.1 Transformational leadership and Team Processes

A studies by (Jung & Sosik, 2002; Wang & Huang, 2009) found that transformational leadership behaviours were positively associated with team cohesion, which is a critical team process that refers to the extent to which change in change in change in team members stick together and remain committed to achieving shared goals. In addition, a meta-analysis of 58 studies by (Lowe et al., 1996) found that transformational leadership was positively related to team performance. Transformational leaders have been found to foster a team culture that encourages open communication, trust, and a shared vision, all of which are critical components of effective team processes (Bass & Avolio, 1993; Wang & Huang, 2009). By inspiring change in change in change in team members to adopt a shared vision, transformational leaders can enhance team commitment and cohesion. They can also promote innovation by encouraging change in change in change in team members to think creatively and challenge the status quo, which can lead to improved team processes and outcomes (Bass & Avolio, 1993).

According to (S. Jha, 2014), transformational leaders can empower change in team members by delegating responsibilities and providing them with the necessary resources and support to achieve their goals. This can help to enhance team self-efficacy, or the belief that the team has the skills and abilities to successfully complete a task. As a result, change in change in change in team members are more likely to take ownership of their work and feel a sense of pride and accomplishment in their achievements, which can further enhance team cohesion and commitment. Additionally, transformational leaders can also influence team processes by promoting a positive and supportive team climate. By creating a culture of respect, fairness, and support, transformational leaders can help to foster a sense of psychological safety among change in change in change in change in team members, which can encourage open communication, collaboration, and constructive feedback. This can help to facilitate effective problem-solving and decision-making processes within the team, which can ultimately lead to improved team outcomes.

H1: *Transformational leadership positively influences Team Processes*

## 3.2 Transformational Leadership and Change in team members

Transformational leadership can cause positive changes in team members' attitudes, perceptions, and motivation towards tasks. A meta-analysis by Judge and Piccolo (2004) of 45 studies found that transformational leadership was positively related to follower attitudes and performance. One study by (Leithwood & Jantzi, 2006) found that transformational leadership was positively related to teachers' motivation to improve their practice. The study by (Avolio et al., 1991) found that transformational leadership was positively related to follower job satisfaction, commitment, and motivation.

Furthermore, transformational leaders have been found to encourage change in team members to adopt a more proactive approach towards tasks and challenges, which can lead to increased motivation and performance (Judge & Piccolo, 2004). Transformational leaders have been found to create a supportive and empowering work environment that enables change in team members to take ownership of their work, feel more engaged, and have a sense of purpose.
H2: *Transformational leadership positively influences Change in team members*

## 3.3 Change in team members and Team Processes
In a study by (Paulsen et al., 2009), it was found team members perceived their team as more cohesive and had a shared sense of identity, they were more likely to engage in cooperative behaviours, which improved team processes. Furthermore, research by (Amabile & Kramer, 2011) showed that change in team members' intrinsic motivation was a key factor in promoting creativity and innovation in teams. When change in team members were intrinsically motivated, they were more likely to engage in behaviours that promoted open communication, risk-taking, and idea-sharing, which can improve team processes and outcomes.
In addition, studies have shown that change in team members' attitudes and perceptions toward tasks can significantly influence their engagement and commitment to achieving team goals. For example, a study by Amabile, (Amabile & Kramer, 2011)found that change in team members who perceived their work as meaningful and important were more likely to be engaged in their work and to persist in the face of challenges, which can positively impact team processes.

A study by (Ilies et al., 2005) found that change in team members' positive attitudes and perceptions towards their work were associated with higher levels of team cohesion and task performance. Similarly, a meta-analysis by (Mullen & Copper, 1994) found that change in team members' motivation to achieve group goals was a critical predictor of team performance. When team members have a positive attitude towards their work, they are more likely to be committed to achieving shared goals and work together more effectively, leading to improved team processes. Moreover, motivated change in change in change in team members tend to be more engaged and involved in the team's work, which can lead to better communication, collaboration, and coordination among change in change in change in team members, all of which are critical components of effective team processes. A study by (Spreitzer et al., 2005) found that team members who perceived their work as meaningful and had a sense of personal control over their work were more motivated and engaged, leading to improved team processes.
H3: *Change in team members positively influence Team Processes*
## 3.4 Change in Change in change in change in team members as a Mediator
A study by (Gooty et al., 2009) found that transformational leadership behaviours were positively associated with employee attitudes towards work. This suggests that transformational leadership can influence employee attitudes, which can have a significant impact on team processes. In addition, a study by (Ehrhart, 2004) found that transformational leadership had a direct effect on employee motivation. This suggests that transformational leadership can influence employee motivation, which can in turn affect team processes. (Schaubroeck et al., 2011) found that the positive effect of transformational leadership on team processes was partially mediated by team members' trust in the leader. A study by (Kim & Yukl, 1995) found that the positive relationship between transformational leadership and team effectiveness was mediated by change in change in change in team members' self-efficacy, which refers to their belief in their ability to perform the task successfully. The study found that transformational leaders can enhance change in change in change in team members' self-efficacy by providing them with positive feedback and support, which, in turn, can improve team processes.

Additionally, (Carmeli et al., 2014) found that the mediating effect of psychological empowerment explained the relationship between transformational leadership and team processes. This suggests that transformational leadership can influence employee attitudes, motivation, and perception, which can then lead to psychological empowerment, ultimately affecting team processes. Overall, change in change in change in change in team members in terms of their attitude, perception, and motivation caused by transformational leadership behaviour will mediate the relationship between transformational leadership and team processes.

H4: *Change in team members mediates the relationship between transformational leadership and Team Processes*

## 4.0 Methods
### 4.1 Questionnaire Design

The scales of transformational leadership measuring thirteen items with higher Cronbach's alphas than the original instrument from (Arif & Mehmood, 2011; Vinger & Cilliers, 2006), were adapted from the work of (Aga et al., 2016). The five-point Likert-type scales were fixed on the extremes of 1 (not important) to 5 (extremely important) to measure the importance of transformational leadership dimension and 1 (strongly disagree) to 5 (strongly agree) was also used to measure the quality of transformational leadership based on the perception of members team. The items for measuring team process measure were adopted from (Kolo, 2017) measuring twelve items of team process measures. The items for measuring change in change in change in change in team members was developed by the researcher after a pilot survey and expert validation. The questionnaire was divided into five sections. Section A is concerned with demographic information; section B sought information on transformational leadership behaviour (idealised influence, intellectual stimulation, inspirational motivation and individualised influence); section C was on the extent to which project manager's transformational leadership behaviour influences project change in change in change in team members' attitude, motivation and perception about task and section D sought information on team processes (communication, cohesion and conflict management). Likert scale format using scale of 1 – Not Important; 2 – Less Important; 3 – Important; 4 – Very Important; 5 – Extremely Important and 1 = Strongly disagree; 2 = Disagree; 3 = Neutral; 4 = Agree; 5 = Strongly Agree was used.

### 4.2 Reliability Research Instruments
Reliability is the degree to which outcomes are consistent and precise to represent the sample frame of a study and if the results of a study can be reproduced using a similar methodology. Using various kinds of processes for gathering data and finding that literature through different quotas extends the reliability of the data and their analysis, as suggested by (Zohrabi, 2013)This study used quantitative methods. Questionnaire of likert scale was the instrument used hence it was imperative to test the internal consistency to ascertain how well they fit the constructs used in the study. Cronbach's Alpha Reliability Coefficient and composite reliability was calculated. A reliability coefficient range between zero (0) and one (1) is adequate (Tavakol & Dennick, 2011). A coefficient of zero means that the instrument has no internal consistency while if it is one means there is internal consistency, (Kothari, 2004) show that a reliable research instrument must have a composite Cronbach's Alpha Reliability Coefficient of at least 0.7 for all indicators used for a study. Table 4.2 shows all the constructs for this study are higher than threshold hence affirming the results of the study.

Table 4.2

| Constructs | Cronbach's Alpha | Number of Items in the Scale |
|---|---|---|
| Change in team members | 0.832 | 13 |
| Team Processes | 0.793 | 13 |
| Transformational Leadership | 0.815 | 12 |

## 4.3 Sample Frame

The target population of this research were construction project team members. Since not all the change in change in change in team members could be studied, a sample of the population was focused on. The study identified the following as those responsible for design and construction of buildings in Nigeria construction industry namely; Architects, Builders, Engineers, Project Managers and Quantity Surveyors. These were recognised as the respondents and main sources of primary data for this research.

## 4.4 Sampling Technique

When it is not possible to collect and analyse data from an entire population in a research study sampling is used (Saunders & Bezzina, 2015)To this end, purposive as well as snowball samplings are more suitable for data collection for this study. Purposive sampling allows the researcher to judge and choose respondents that were working in construction project teams and thus were most likely to answer the research questions and to satisfy the objectives of the research (Saunders & Bezzina, 2015). Self-administered questionnaires are usually answered by the participants themselves, while researcher-administered questionnaires answers were recorded by the researcher (Saunders & Bezzina, 2015). Self-administered questionnaires were adopted for this research because it does not allow the researcher to influence the respondent to cause bias (MacIntosh & O'Gorman, 2015).

## 4.5 Sample Size

Since the population of this study (change in change in change in team members of construction projects) is unknown the procedure developed by Cochran (1977) see (Kotrlik & Higgins, 2001), for determining the sample of an infinite population was adopted. The formula is given as follows:

$$N = \frac{z^2 \times p \times q}{e^2}$$

Where: N = Sample Size, z = selected critical value of desired confidence level (1.95), p = estimated proportion of an attribute that is present in the population (0.5), q = 1 - p and e = desired level of precision (0.08).
Therefore N = 149.
40 percent of N was added to cover for non-return rate which is 60 hence N = 209

## 4.6 Data Analysis Technique

This study used both descriptive and inferential statistical analysis to test the study hypothesis. Non parametric descriptive analysis was also used to determine central tendency and measures of dispersion. The arithmetic mean was measure for central tendency while standard deviation was the measure of dispersion. Partial Least Square Path Modeling (SMART PLS 3) was used to analyse the effect of mediating variable on the relationship between independent variables and the dependent variable. The choice of PLS analysis method was chosen since it was used to measure and explain the link between independent variables and dependent variable, the relationship among the mediating variables. Beta Coefficient was used to measure the effect of the mediating variable on the relationship between the independent and dependent variables under study. In testing the hypothesis PLS's (R2), was used. Additionally, with the aid of PLS path modeling, this study aims to assess transformation leadership with the mediating role of change in change in change in team members in the relationship between transformational leadership and team process. This Study utilised PLS path modeling to evaluate the proposed model in Nigerian construction industry.

In measuring the hypothesised model, Smart PLS 3 was utilised to determine the parameters of the model. In this case, PLS path modelling was applied with a path-weighting scheme for inside approximation (Ramayah et al., 2018; Tenenhaus et al., 2005; Wetzels et al., 2009). Then, nonparametric bootstrapping with 10,000 replications to obtain the standard estimate errors was carried out (Ramayah et al., 2018). Method of repetitive indicators was conducted to assess the higher order latent variable, in accordance with (Lohmöller & Lohmöller, 1989; Tibshirani & Efron, 1993).

Endogeneity issue can be as a result of two situations in a research model: (1) it may occur if any bidirectional relationship is expected among some of the concepts in the model ((Abdallah et al., 2015) and (2) when there are limited disregarded variables that could also incorporated in the controlled model such that the effect of "X on Y" cannot be understood as a result of missing causes (Antonakis, 2001). However, limited researchers deal with endogeneity in PLS model (Lovaglio & Vittadini, 2013). On the basis of this scarce studies and the assertion on possible links amongst predictors and outcomes within the explanatory equations of the PLS model that can be influenced by unmodeled constituents in the predictor blocks, the present study deleted those additional factors from the predictor blocks. Thus, additional factors have been fully integrated into the hypothesised model that the endogeneity issue has been removed (Lovaglio & Vittadini, 2013).

## 5.0 Results

### 5.1 Demographic of the Respondents

The results presented in Table 5.1, details the demographics of a study on construction project change in change in change in team members. The study identified five portfolios: Architects, Builders, Engineers, Project Managers, and Quantity Surveyors. The data shows that the highest proportion of change in change in change in team members were Architects, followed by Quantity Surveyors, Builders, and Engineers. The majority of respondents were between 31-40 years old and held a bachelor's degree. Regarding experience, most had 5-10 years of experience working in design teams, and the majority worked for contracting or consultancy organisations in medium or large projects, mainly in residential and institutional buildings. Only 3.5% of team leadership was provided by project managers.

Table 5.1 Demographics of the respondents

| Respondents' Distribution | Frequency | Percentage (%) |
| --- | --- | --- |

| Profession: | | |
|---|---|---|
| Project Manager | 5 | 3.5 |
| Engineer | 27 | 19.5 |
| Architect | 41 | 29.1 |
| Builder | 33 | 23.4 |
| Quantity Surveyor | 35 | 24.8 |
| **Age:** | | |
| <30 | 46 | 32.6 |
| 30-40 | 56 | 39.7 |
| 41-50 | 26 | 18.4 |
| 50> | 13 | 9.2 |
| **Qualification:** | | |
| ND/HND | 9 | 6.4 |
| BSc | 8 | 60.3 |
| MSc | 43 | 9.1 |
| PhD | 2 | 1.4 |
| **Years of experience:** | | |
| <5 years | 47 | 33.3 |
| 5-10 | 50 | 35.5 |
| 10-15 | 34 | 24.1 |
| 15> | 10 | 7.1 |
| **Project Leader:** | | |
| Project Manager | 5 | 3.5 |
| Engineer | 27 | 19.2 |
| Architect | 41 | 29.1 |
| Builder | 33 | 23.4 |
| Quantity Surveyor | 35 | 24.8 |
| **Organisation:** | | |
| Client | 25 | 17.7 |
| Consultant | 58 | 41.1 |
| Contracting | 58 | 41.1 |
| **Project Size:** | | |
| Small | 10 | 7.1 |
| Medium | 65 | 46.1 |
| Large | 66 | 46.8 |
| **Project Type:** | | |
| Residential | 61 | 43.3 |
| Commercial | 32 | 22.7 |
| Institutional | 48 | 34.0 |
| **Project Status:** | | |
| Completed | 77 | 54.6 |
| On-going | 64 | 45.4 |

## 5.2 Descriptive Statistics

The first column lists in table 5.2 is the variable names, the second column provides the sample size, and the third and fourth columns display the minimum and maximum values used for calculating the mean from data collected. A five-point Likert scale ranging from one to five was used for all three variables. The independent variable, which is transformational leadership, has a mean of 3.7589 and a standard deviation of 0.53319. The dependent variable, team processes, has mean and standard deviation values of 3.8227 and 0.60100, respectively. The mediator variable in this study, change in change in change in change in team members, has a mean of 3.8156 and a standard deviation of 0.67191.

Table 5.2: Descriptive Statistics

| Variables | Sample Size | Minimum | Maximum | Mean | Std. Deviation |
|---|---|---|---|---|---|
| Transformational Leadership | 141 | 3.00 | 5.00 | 3.7589 | 0.53319 |
| Change in change in change in team members | 141 | 2.00 | 7.00 | 3.8156 | 0.67191 |
| Team Processes | 141 | 2.00 | 6.00 | 3.8227 | 0.60100 |

## 5.3 Correlation analysis

As proposed, there are significant and positive correlations among transformational leadership, change in change in change in team members, and team processes. Specifically, transformational leadership was found to be positively and significantly correlated with change in change in change in team members (r = .294**, p < 001) and with team processes (r = .178**, p < 001). Additionally, team processes and change in change in change in team members were significantly correlated (r = .308**, p < 001). The study's proposed model was tested using Partial Least Squares (SMART PLS 3) software, and the results are displayed in Table 5.4.

Table 5.3 Correlations Analysis

| Variables | 1 | 2 | 3 |
|---|---|---|---|
| 1 Transformational Leadership | - | | |
| 2 Change in change in change in team members | .294** | - | |
| 3 Team Processes | .178** | .308** | - |

Correlation is significant at the 0.01 level (2-tailed)

## 5.4 Overall Path Model

The structural equation model is considered as the second major process of structural equation modeling analysis. Once validation process of the measurement model is confirmed, then representation of the structural model can be established by identifying the relationships between the concepts. The structural model provides details on the relationships between the constructs (Lowry & Gaskin, 2014). It shows the specific details of the relationship among the independent or exogenous and dependent or endogenous variables (Sarstedt et al., 2020). Evaluation of the structural model focus firstly on the overall model fit, followed by the size, direction and significance of the proposed parameter estimates, as displayed by the one- headed arrows in the path diagrams (Hair et al., 2014). The final part included the confirmation process of the structural model of the study, which was established on the projected relationship among the identified and assessed variables. In the present study, the structural model was supposed to test the research hypothesizes, through Partial Least Square (SMART PLS 3) Path Modeling technique and bootstrapping with 10000 replications.

In order to specify the research hypotheses targeted in chapter one of this work, a research structural model was developed in this study. The research structural model is proposed to test four hypotheses related to direct effects from transformational leadership, change in change in change in team members on team processes and transformational leadership on change in change in change in team members. The study also examined the mediation effects of change in change in change in team members on the relationship between transformational leadership and team processes. Figure 1 illustrates the hypothesised direct and mediation effects in the research Path Modeling.

Partial Least Square Path Modeling is a data analytic approach generally used to assess patterns of relationships that exist among variables (Lowry & Gaskin, 2014). The latent constructs in individual CFA models were all measured by several multi-item scales. The inclusion of all items and relative errors in the measurement and structural models leads to a complex and non-stable model because too many parameters need to be estimated. Thus, to overcome this problem, this research utilised parcels as indicators of latent constructs in individual CFA models. Parcels are aggregations (sums or averages) of several individual items. Using parcels as indicators of latent construct commonly have better reliability as compared with the single items (Coffman & MacCallum, 2005). As the result of using item-parcelling procedure, the latent constructs in individual CFA models of (transformational leadership, change in change in change in team members and team processes) were converted into observed variables so that they could easily construct the overall measurement and structural model and reduce the model complexity. Confirmatory factor analysis was used to examine the overall measurement model. The model comprises all of the first and second order constructs proposed in this study. Figure 5.4 depicts the overall Path model.

Figure 5.4: Overall Path Model

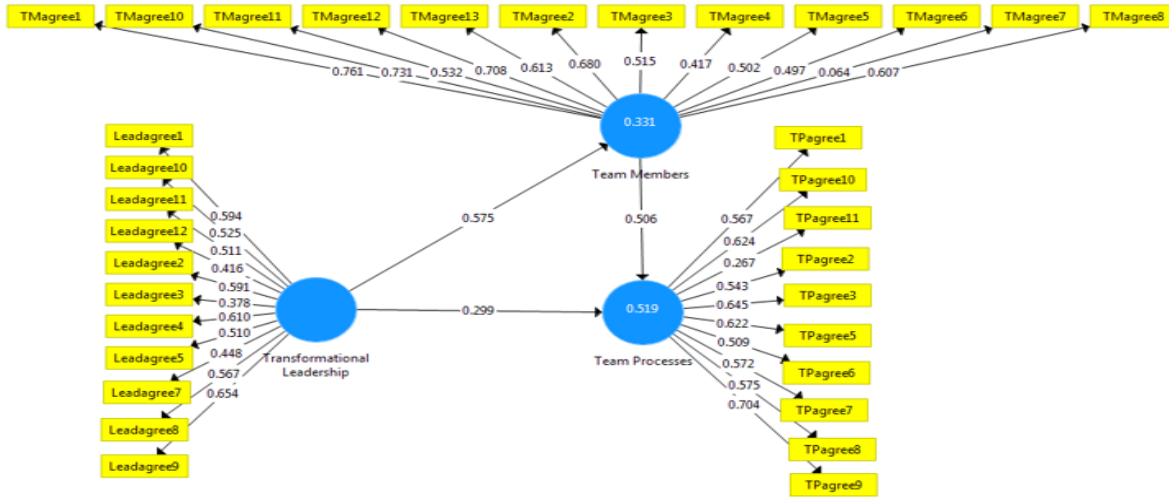

### 5.4.1 Reliability and Convergent Validity

Table 5.4.1 represents the result of Cronbach's alpha and convergent validity for the Overall CFA model. The results of assessing the standardized factor loadings of the model's items indicated that the initial standardised factor loadings of some items were below 0.6, ranged from 0.267 to 0.594 but other item were above 0.6. Once the uni-dimensionality of the constructs was achieved, each of the constructs was assessed for their reliability. Reliability is assessed using average variance extracted (AVE), composite reliability (CR) and Cronbach's alpha. Table 5.5 shows that the AVE values were 0.285, 0.337, and 0.329 for transformational leadership, change in change in change in team members and team processes respectively. All of these values were below the cut-off of 0.5 as suggested by (Lowry & Gaskin, 2014) however, Average Variance Extracted (AVE) is strict measure of reliability, and the other two measures (Cronbach's alpha and composite reliability) were used to assess the reliability of the constructs. The Cronbach's Alpha values were 0.817, 0.855 and 0.825 for transformational leadership, change in change in change in team members and team processes respectively. These values were all above the threshold of 0.7 as suggested by (Charter, 1999). The composite reliability values were, 0.811, 0.847, and 0.825 for transformational leadership, change in change in change in team members and team processes respectively. These values exceeded the recommended value of 0.6 for all constructs as recommended by (Lowry & Gaskin, 2014).

Table 5.4.1 Reliability and Convergent Validity

| Construct | Items | Loadings | Alpha | CR | AVE |
|---|---|---|---|---|---|
| Transformational Leadership | Leadagree1 | 0.594 | 0.817 | 0.811 | 0.285 |
| | Leadagree10 | 0.525 | | | |
| | Leadagree11 | 0.511 | | | |
| | Leadagree12 | 0.416 | | | |
| | Leadagree2 | 0.591 | | | |
| | Leadagree3 | 0.378 | | | |
| | Leadagree4 | 0.61 | | | |
| | Leadagree5 | 0.51 | | | |
| | Leadagree7 | 0.448 | | | |
| | Leadagree8 | 0.567 | | | |
| | Leadagree9 | 0.654 | | | |
| Change in change in change in team members | TMagree1 | 0.761 | 0.855 | 0.847 | 0.337 |
| | TMagree10 | 0.731 | | | |
| | TMagree11 | 0.532 | | | |
| | TMagree12 | 0.708 | | | |
|  | TMagree13 | 0.613 |  |  |  |
|  | TMagree2 | 0.68 |  |  |  |
|  | TMagree3 | 0.515 |  |  |  |
|  | TMagree4 | 0.417 |  |  |  |
|  | TMagree5 | 0.502 |  |  |  |
|  | TMagree6 | 0.497 |  |  |  |
|  | TMagree7 | 0.064 |  |  |  |
|  | TMagree8 | 0.607 |  |  |  |
| Team Processes | TPagree1 | 0.567 | 0.825 | 0.825 | 0.329 |
|  | TPagree10 | 0.624 |  |  |  |
|  | TPagree11 | 0.267 |  |  |  |
|  | TPagree2 | 0.543 |  |  |  |
|  | TPagree3 | 0.645 |  |  |  |
|  | TPagree5 | 0.622 |  |  |  |
|  | TPagree6 | 0.509 |  |  |  |
|  | TPagree7 | 0.572 |  |  |  |
|  | TPagree8 | 0.575 |  |  |  |
|  | TPagree9 | 0.704 |  |  |  |

AVE = Average Variance Extracted = (summation of the square of the factor loadings)/{(summation of the square of the factor loadings) + (summation of the error variances)}; CR = Composite reliability = (square of the summation of the factor loadings)/{(square of the summation of the factor loadings) + (square of the summation of the error variances)} and Alpha = Cronbach's Alpha.

### 5.4.2 Discriminant Validity

Heterotrait-Monotrait Ration (HTMT) criteria is a measure of discriminant validity that says actually whether this are the same or differently in factors, if the value are less than one (<1) (Voorhees et al., 2016) that indicate that this are different factors, notice the more lower they are, the more different they are, so in this case there is discriminant validity because the (HTMT) values 0.666, 0.555, and 0.584 are less than 1 (see below; table 5.4.2).

Table 5.4.2 Heterotrait-Monotrait Ration (HTMT)

|  | Change in change in change in team members | Team Processes | Transformational Leadership |
|---|---|---|---|
| Change in change in change in team members |  |  |  |
| Team Processes | 0.666 |  |  |
| Transformational Leadership | 0.555 | 0.584 |  |

### 5.5 Hypothesis Testing

**Hypothesis 1**, states that transformational leadership positively influences team processes. Results of the Partial Least Square Path Modelling (SMART PLS 3) analysis are depicted in Table 5.5 The result indicates that transformational leadership has a significant and positive relationship with team processes ($\beta = 0.630$, $P < 0.000$) and the r2 value (0.397) satisfies the requirement for the 0.30 cut off value as recommended by (Lowry & Gaskin, 2014) which uniquely explains 39.7 percent of the variance in team processes. Therefore, Hypothesis 1 is supported.

**Hypothesis 2** proposes that transformational leadership is positively related to change in change in change in team members. The results of Partial Least Square Path Modelling (SMART PLS 3)

Table 5.5 indicate a strong and highly significant relationship between transformational leadership and change in change in change in team members do exist (β = 0.633, P < 0.000) and the r2 value (0.401) satisfies the requirement for the 0.30 cut off value as recommended by (Lowry & Gaskin, 2014) which uniquely contributed 40.1 percent of the variance in change in change in change in team members. Hypothesis 2 is therefore supported.

**Hypothesis 3** states that change in change in change in team members are positively related to team processes. The results of Partial Least Square Path Modelling (SMART PLS 3) Table 5.5 indicate that change in change in change in team members with r2 value (0.479) satisfies the requirement for the 0.30 cut off value as recommended by (Lowry & Gaskin, 2014) which uniquely contributed 47.9 percent of the variance in team processes and show a strong and highly significant relationship between change in change in change in team members and team processes (β = 0.692, P < 0.001). Hypothesis 3 is therefore supported.

Table 5.5 Standardized Direct path coefficients of the hypothesized model

|  | (O) | (M) | (STDEV) | (\|O/STDEV\|) | P Values |
|---|---|---|---|---|---|
| Change in change in change in team members -> Team Processes | 0.692 | 0.741 | 0.047 | 14.709 | 0. 000 |
| Transformational Leadership -> Change in change in change in team members | 0.633 | 0.680 | 0.067 | 9.492 | 0. 000 |
| Transformational Leadership -> Team Processes | 0.630 | 0.678 | 0.060 | 10.450 | 0.031 |

O= Original Sample, M= Sample Mean, STDEV= Standard Deviation

**5.6 Mediating Effects**

Table 5.6 presents the results of investigating the mediating effect of change in change in change in team members on the relationship between transformational leadership and team processes (Hypothesis 4). Before conducting the analysis, the conditions for mediation analysis were examined. The first condition was met, as the results indicate a significant positive influence of transformational leadership on team processes (β = 0.630, P < 0.000), demonstrating a correlation between the independent and dependent variables. The second condition was also satisfied, as the results in Table 5.6 reveal a significant positive relationship between transformational leadership and change in change in change in team members (β = 0.633, P < 0.000), providing evidence for a significant relationship between the independent variable and the mediator variable. The third condition was confirmed by the results in Table 5.6, which show a significant relationship between change in change in change in team members, the mediator variable, and team processes (β = 0.692, P < 0.001). When the mediator (change in change in change in team members) was introduced in the model, the direct relationship between the independent variable (transformational leadership) and dependent variable (team processes) was substantially reduced, from (β = 0.630, P < 0.000, as seen in Table 5.6) to (β = 0.299, P < 0.000, as seen in Table 5.6.1), indicating the presence of a mediating effect (Lowry & Gaskin, 2014).

Table: 5.6 indirect path coefficients of the hypothesized model with mediator

|  | (O) | (M) | (STDEV) | (O/STDEV) | P Values |
|---|---|---|---|---|---|
| Change in change in change in team members -> Team Processes | 0.506 | 0.530 | 0.102 | 4.979 | 0. 000 |
| Transformational Leadership -> Change in change in change in team members | 0.575 | 0.604 | 0.084 | 6.815 | 0. 000 |
| Transformational Leadership -> Team Processes | 0.299 | 0.293 | 0.113 | 2.645 | 0.008 |

5.6.1 Specific Indirect path coefficients of the hypothesized model

|  | (O) | (M) | (STDEV) | (O/STDEV) | P Values |
|---|---|---|---|---|---|
| Transformational Leadership -> Change in change in change in team members -> Team Processes | 0.291 | 0.320 | 0.077 | 3.763 | 0.000 |

Note: n=141; Bootstrap sample size=10000, BC 95% CI= Bootstrap confidence Intervals
*p<.05, **p<.01, ***p<.00

In order to confirm the mediating effect, the specific indirect effect of a × b in equation (1) had to be significant. As shown in Table 5.6.2, the Z statistic procedure proposed by (Sobel, 1982) was computed and found to be significant at p < 0.000. The Z value 4.02 is greater than the t-value of >1.96 (for 2-tailed), which is equivalent to p < 0.05. Therefore, Hypothesis H4 was accepted, indicating that there is an indirect influence from transformational leadership through change in change in team members on team processes. It is worth noting that there are three main types of the Sobel test, including adding the third denominator part (Aroian, 1947) as suggested by (Preacher & Leonardelli, 2001), subtracting the third denominator part (Goodman, 1960), and not considering the denominator. Before the Sobel test was applied to parameter estimates obtained from modeling, the procedures given by (MacKinnon et al., 2002) and (Krull & MacKinnon, 1999) were considered. The Z value (Krull & MacKinnon, 1999) is mathematically defined as follows:

$$Z - Value = \frac{A \times B}{\sqrt{B^2 \times SEA^2 + A^2 \times SEB^2}} \quad \ldots\ldots\ldots\ldots\text{equation (1)}$$

Where: A= Transformational Leadership → Change in change in team members, B= Change in change in team members → Team Processes, SEA = Standard Error of A and SEB = Standard Error of B (Table 5.6 above)

$$Z - Value = \frac{0.575 \times 0.506}{\sqrt{(0.506)^2 \times (0.084)^2 + (0.575)^2 \times (0.102)^2}}$$

**Z-Value = 4.02**

Table 5.6.2: Sobel Test Result

| Sobel Test | 1-tailed | 2-tailed | Standard deviation | P-values |
|---|---|---|---|---|
| 4.01686415 | 0.00002949 | 0.00005898 | 0.07243212 | 0.00005898 |

As shown in Figure 5.4, there was a significant influence from transformational leadership on change in team members (0.575, p <0.01) and change in team members on team processes (0.506, p <0.01). The Z value score also greater than 1.96 (p <0.05); thus, the ultimate result approves the mediating effect of change in team members on the relationship between transformational leadership and team processes, which in turn indicates that it has an indirect effect on team processes. To calculate the component of the indirect effect, the variance accounted for (VAF) value was considered, which is the ratio of indirect effect to total effect. The VAF value depicts that 49.3% of the total effect of transformational leadership on team processes was as the result of the indirect effect of change in team members. $\frac{A \times B}{A \times B + C}$

Where: A= Transformational Leadership → Change in team members, B= Change in team members → Team Processes and C= Transformational Leadership → Team Processes (table 5.6)

$$VAF = \frac{0.575 \times 0.506}{[(0.575 \times 0.506) + 0.299]} \quad \textbf{VAF = 0.493 = 49.3\%}$$

## 5.7 Discussion

The study's first hypothesis, which posited that Transformational Leadership would enhance the likelihood of positive Team Processes, was strongly supported by the results. The study aimed to explore the connection between transformational leadership and team processes by examining the mediating influence of changes in change in team members. The findings showed a positive correlation between a project manager's transformational leadership style and team processes, indicating that the leader's influence is essential in fostering effective team processes. A transformational project manager motivates and inspires change in team members to adopt a general notion of team processes that involve effective communication, cohesion, and conflict management. This result underscores the importance of project managers' leadership styles, an area that previous project management literature has neglected, as highlighted by (Turner & Müller, 2005).

Furthermore, the study found that the influence of transformational leadership on changes in change in team members positively impacts team processes, which supports the findings of (Yang et al., 2011). Transformational leadership is one of the most critical styles of leadership as it constantly seeks significant changes in teams and involves change in team members in decision-making. Transformational leaders articulate a vision that is attractive, desirable, and achievable to change in team members. They make valuable contributions to the success of teams by motivating and encouraging change in team members and creating a supportive work environment that fosters coordination and assistance.

Transformational leadership techniques such as increasing motivation and task performance also enhance team cohesion. Such leaders encourage team identity and encourage change in team members to transcend their self-identity. In summary, the findings of this study demonstrate that transformational leadership is a crucial component of effective team processes and that project managers should strive to develop and apply this leadership style to foster positive outcomes. Transformational leaders inspire their change in team members to work in the best interest of the team by serving as role models and providing psychological ownership to their followers. By identifying their followers' needs and strengths, these leaders guide their behaviour towards teamwork, resulting in the successful completion of projects. The relationship between transformational leaders and their followers is based on strong ethical and moral principles, which increase motivation for both parties, enabling them to collaborate in the best interest of the team. Through articulating an attractive vision, transformational leaders can change their change in team members' attitudes, perceptions, and behaviours by directing them towards the common goal of the team. Unlike other forms of leadership, such as transactional leadership, which focus on economic and transactional exchanges, transformational leaders are able to change the personality of their change in team members by presenting challenging tasks and an appealing vision. Transformational leaders have an idealized influence and serve as role models for their followers, directing their efforts towards the team's success. They also provide an encouraging environment that helps achieve various desirable outcomes, such as improved job performance, creativity, and organisational citizenship behaviour among employees. These behaviours contribute to the success of project-based organisations, helping them realize their goals in an effective and efficient manner.

Transformational leadership is characterized by strong moral values and concern for change in team members, which results in high levels of trust, respect, and admiration from change in team members towards their leaders. This leadership style involves changing the ways of working by providing new directions for change in team members to perform their roles for the benefit of the project and achieving their own goals. Under transformational leaders, followers are willing to perform beyond expectations and their formal roles, using their full energy and putting all their efforts into their work. They sacrifice their self-interest for the collective interest of the team, increasing the chances of success.

Individual consideration is a key characteristic of transformational leadership, with leaders showing strong consideration for follower needs and encouraging new ideas and methods from their followers through intellectual stimulation. Transformational leaders also challenge the status quo, encouraging their followers to come up with new ideas and presenting ways of doing things in new ways. By providing a supportive environment, change in team members are encouraged to engage in creative work performance, and leaders provide incremental and psychological support when needed. Transformational leadership has four highly effective components. First, leaders serve as role models, exhibiting idealized influence. Second, they motivate followers through inspirational motivation. Third, they show strong concern for the needs, feelings, and emotions of followers through individualized consideration. Finally, they increase the intellectual level and knowledge, skills, and abilities of their followers.

This positive leadership style is highly influential in project-based organisations, triggering intrinsic motivation, developing employee skills, increasing moral standards, initiating changes, increasing maturity levels, creating a supportive climate for teamwork, promoting sacrifice of self-interest for team goals, promoting coordination and cohesion, being consistent with words and actions, coaching subordinates, transforming lives, and considering the input of followers through task significance.

The fourth hypothesis of the current study was confirmed as significant, as the basic assumptions of mediation were satisfied before testing the mediating role of change in team members. This study suggests that changes in change in team members' attitudes, motivation, and perception towards work can result in a highly empowered and dedicated project team. Improving these factors can also enhance change in team members' knowledge of team processes, interpersonal communication, and problem-solving skills, leading to improved team processes. Additionally, the study found that changes in change in team members mediate the relationship between a project manager's transformational leadership and team processes. This is the first study to recognize the mediating role of changes in change in team members in the connection between transformational leadership and team processes. This finding contributes to understanding how transformational leadership impacts team processes and shows that project managers who demonstrate transformational leadership behaviour are more likely to create changes in change in team members' attitudes, motivation, and perception towards work that support the achievement of team processes.

Transformational leaders are known for their individualized consideration for each of their followers and their ability to show respect and concern for their followers' needs. They serve as role models for morality, ethics, justice, and fairness, both on and off the job, and support their followers in effectively resolving problems. These leaders also develop their followers' personality, knowledge, skills, and abilities, leading to positive attitudes such as trust, loyalty, respect, team identification, and commitment to the team and leader. These attitudes lead to favorable employee behaviours such as organisational citizenship behaviour, personal initiative, voice behaviours, low turnover intention, and teamwork. All of these attitudes and behaviours are based on the changes transformational leaders create in their change in team members, which promote team processes. Transformational leaders pay attention to the needs of their change in team members, and the change in team members respond by putting in their complete effort and enthusiasm, which boosts team processes such as communication, cohesion, and conflict management. High-quality relationships based on trust, loyalty, respect, and affiliation between transformational leaders and their followers lead to improved team processes such as communication, cohesion, and conflict management.

The influence of transformational leadership on change in team members leads to changes in their attitudes, motivation, and perception towards tasks, which enhance team processes. This change is highly required in project-based organisations in order to foster team processes. Such changes lead to desirable behaviours such as affective commitment with team leaders, increased job satisfaction and supervisor perception, and perception in terms of rewards and procedures, which further improve team processes. The influence of transformational leadership on change in team members also leads to increased innovative behaviour and creativity in teams, as change in team members feel comfortable with their leaders and are more likely to come up with interesting ideas. Such relationships are building blocks for team processes and lead to improved conflict management, cohesion, and communication among change in team members.

**5.8 Conclusion and Future Research**

In conclusion, this study has provided evidence that supports the hypothesis that transformational leadership enhances positive team processes. The findings demonstrate that transformational leadership is a crucial component of effective team processes, and project managers should strive to develop and apply this leadership style to foster positive outcomes. Transformational leaders inspire their change in team members to work in the best interest of the team by serving as role models and providing psychological ownership to their followers. Furthermore, this study recognised the mediating role of changes in change in team members in the connection between transformational leadership and team processes. This finding contributes to understanding how transformational leadership impacts team processes and shows that project managers who demonstrate transformational leadership behaviour are more likely to create changes in change in team members' attitudes, motivation, and perception towards work that support the achievement of team processes. Overall, the study underscores the importance of transformational leadership in project-based organisations and highlights the significant impact that transformational leaders can have on team processes and outcomes.

Future research could investigate how transformational leadership affects different types of teams and team processes. It would be interesting to examine whether the impact of transformational leadership varies based on the size, type, or complexity of the team, and whether the effect is more significant for certain team processes, such as communication or conflict management. Future research could investigate the role of team members' personality traits, values, and attitudes in mediating the relationship between transformational leadership and team processes. It would be interesting to examine whether some team members are more receptive to transformational leadership than others and how their individual characteristics influence the effectiveness of this leadership style. Finally, future research could explore the impact of training and development programs on project managers' ability to demonstrate transformational leadership behaviours. It would be valuable to investigate how project managers can be trained and developed to enhance their leadership skills and foster more effective team processes.